\documentclass[twocolumn,showpacs,aps,pra,letterpaper,nofootinbib,raggedbottom,nobalancelastpage]{revtex4}
\usepackage{graphicx}
\usepackage{amsmath}
\usepackage{color}

\begin{document}
\title{Generation of the superposition of mesoscopic states of nano-mechanical resonator by a single two-level system}
\date{\today}
\author{Manoj Das, J. K. Verma and P. K. Pathak}
\address{School of Basic Sciences,
Indian Institute of Technology Mandi, Kamand, H.P. 175005, India}
\begin{abstract}
We propose measurement based conditional generation of superposition of motional states of nanomechanical resonator. We consider a two level quantum mechanical system coupled with nanomechanical resonator through phonon exchange. An interaction, which produces shifts in the state of nanomechanical resonator depending on the state of the qubit, is realized by driving qubit through two resonant lasers. The measurement of the state of quantum mechanical system produces superposition states of nanomechanical resonator. We show that the quantum interference between states in superposition may lead to arbitrary large displacement in resonator. We also discuss decoherence of generated states using Wigner function.
\end{abstract}
\maketitle
\section{Introduction}
Nanomechanical resonators (NMR) have wide range of applications which includes ultra-sensitive mass detection \cite{mass}, force detection \cite{force1,force2}, imaging and lithography \cite{imaging}. Recently it has been possible to cool fundamental mode of NMR to its ground state \cite{cooling,Oconnell}and creating single phonon excitation in a controlled manner\cite{Oconnell}. Clearly, it is a milestone in demonstrating quantum mechanical behaviour of a mesoscopic system consisting of billions of atoms,  which can lead to applications such as quantum limited measurements\cite{qlimited}, generating nonclassical motional states\cite{cat}, realization of hybrid quantum mechanical systems for quantum information processing\cite{cavitycoupled, bothcavity,cpb,jj,nv,nv2}. NMR not only provide high quality factor but also provide ease to couple various quantum mechanical systems through magnetic, electrical or optical interactions. Such developments have opened up important possibilities to encode quantum information into mechanical state and mediating interaction between different types of quantum systems. Two types of hybrid system have been realized by coupling NMR to either a cavity where electromagnetic field directly couples through radiation pressure\cite{cavitycoupled,bothcavity} or by coupling with a qubit where states of qubits can be manipulated by external fields\cite{cpb,jj,nv,nv2,nanolett}. In the case of cavity coupled NMR various phenomenon such as optomechanical induced transparancy\cite{meit}, single photon blockade\cite{blockade}, and generating micro-macro entangled states\cite{micro-macro} have been realized.

Due to versatile nature, NMR has been efficiently coupled to various types of qubits such as superconducting qubit\cite{cpb,jj}, quantum dots\cite{nanolett}, cold atoms\cite{coldatom}, nitrogen vacancy (NV) centers\cite{nv,nv2} which provides mechanical analog of cavity QED. These systems have vital importance particularly for controlling and generating nonclassical states of NMR.
There have been proposals for cooling\cite{imamoglu} or lasing\cite{phlaser} NMR by coupling with a two level quantum system. A quantum dot (QD)\cite{nanolett} or a NV center\cite{nv2} embedded in nanowire naturally couples through phonon mode by strain mediated interaction. The effects of such coupling have been witnessed in florescence spectra from quantum dot or NV center driven by an external field. QD or NV center embedded in nanowire also provides very high collection efficiency  for emitted photons which has been utilized in realization of high efficiency single photon sources\cite{sps}.

In this paper we present a scheme for controlling state of NMR by externally deriving a coupled two level system and measuring the state of the system.Our paper is organized as follows. In Sect.~\ref{model}, we present our model for realization of effective interaction.
Sect.~\ref{mesoscopic} presents method for generating superposition of mesoscopic states of NMR.
In Sect.~\ref{Sec:Decoherence} we discuss effects of decoherence.
Finally, we present possibilities for experimental realization in current scenario and conclude in Sect.~\ref{Sec:Conclusions}.

\section{Model for realization of effective Hamiltonian}
\label{model}
We consider a qubit coupled with NMR, the qubit is driven by two external resonant fields, the fields have phase difference of $\pi/2$. The Hamiltonian of the system in rotating frame is given by
\begin{eqnarray}
H=\hbar\omega a^{\dagger}a+\hbar\Omega_1(\sigma^++\sigma^-)+i\hbar\Omega_2(\sigma^+-\sigma^-)\nonumber\\
+\hbar g |e\rangle\langle e|(a+a^{\dagger});
\label{ham1}
\end{eqnarray}
where $\omega$, $a$ ($a^{\dagger}$), $\Omega_1$ ($\Omega_2$), $g$, $\sigma^-$, and $\sigma^+$ are frequency of NMR mode, annihilation (creation) operator for phonon field, coupling constants for qubit with first (second) external field, coupling constant for qubit with NMR mode, lowering and raising operator for qubit, respectively. We notice that similar Hamiltonian has been considered earlier by Freedhoff and Ficek \cite{doublydriven1} in the context of modification in Mollow's sidebands and the results have been recently verified by He et al\cite{doublydriven}.
Under canonical transformation\cite{imamoglu} $H\rightarrow e^sHe^{-s}$, with $s=\eta|e\rangle\langle e|(a^{\dagger}-a)$; where $\eta=g/\omega$, above Hamiltonian takes the form
\begin{eqnarray}
H=\hbar\omega a^{\dagger}a+\hbar\Omega_1[\sigma^+e^{\eta(a^{\dagger}-a)}+\sigma^-e^{-\eta(a^{\dagger}-a)}]\nonumber\\
+i\hbar\Omega_2[\sigma^+e^{\eta(a^{\dagger}-a)}-\sigma^-e^{-\eta(a^{\dagger}-a)}].
\label{ham2}
\end{eqnarray}
 For $\eta\ll1$ this canonical transformation has been widely utilized in ion trap for deriving effective interaction. For QD and superconducting qubits ($g\sim 100KHz$) coupled with NMR having $\omega\sim 1GHz$ and for NV centers ($g\sim 1KHz$) coupled with NMR having $\omega\sim1MHz$; $\eta\sim 10^{-3}$. Therefore we consider the terms up to second order in $\eta$ and also consider that the qubit is strongly driven by one of the field, say $\Omega_1\gg g, \omega, \Omega_2, \Gamma$. We further rewrite above Hamiltonian in the interaction picture in which the interaction with stronger field has been diagonalized. In this picture the state of the qubit  and the Hamiltonian $H$ is transformed to
\begin{eqnarray}
|\psi\rangle=e^{iht}|\psi\rangle,~~H=e^{iht}He^{-iht}; h=\Omega_1(\sigma^++\sigma^-)
\end{eqnarray}
The qubit spin operators $\sigma^{\pm} $ transform as
\begin{eqnarray}
e^{iht}\sigma^{\pm}e^{-iht}=\sigma^{\pm}\cos^2\Omega_1t+\sigma^{\mp}\sin^2\Omega_1t\mp i\sigma_z\sin2\Omega_1t
\label{sigma}
\end{eqnarray}
We consider that qubit is driven strongly such that $\Omega_1$ is very large therefore we can neglect highly oscillating terms in Eq(\ref{sigma}), i.e. $\sin2\Omega_1t\approx0$.  Similar approximation has been used earlier extensively in various contexts\cite{intense}. Under this approximation the effective Hamiltonian becomes
\begin{eqnarray}
\bar{H}_{eff}=\hbar\omega a^{\dagger}a+i\hbar\Omega_2\eta(\sigma^++\sigma^-)(a^{\dagger}-a)\nonumber\\
+\frac{\hbar\Omega_1\eta^2}{2}(\sigma^++\sigma^-)(a^{\dagger}-a)^2
\label{heffbar}
\end{eqnarray}
We note that $\bar{H}_{eff}$ commutes with $h$, thus changing to original interaction picture and neglecting two phonon transitions the effective hamiltonian becomes
\begin{eqnarray}
H_{eff}=\hbar\omega a^{\dagger}a+\hbar\Omega^{\prime}_1(\sigma^++\sigma^-)\nonumber\\
+i\hbar\Omega_2\eta(\sigma^++\sigma^-)(a^{\dagger}-a)-\hbar\Omega_1\eta^2(\sigma^++\sigma^-)a^{\dagger}a
\label{heffbar}
\end{eqnarray}
where $\Omega^{\prime}_1=\Omega_1(1-\eta^2/2)$. The Hamiltonian $H_{eff}$ can be diagonalized in the basis of QD, $|\pm\rangle=(|e\rangle\pm|g\rangle)/\sqrt{2}$, as follows
\begin{eqnarray}
H_{eff}\approx\hbar\Omega^{\prime}_1\lambda+\hbar(\omega-\Omega_1\eta^2\lambda)A^{\dagger}A
\end{eqnarray}
where $A=a+i\Omega_2\eta\lambda/\omega$, and $\lambda=\pm1$ corresponding to the states $|\pm\rangle$. The first term gives the change in phase, the second term gives change in phase as well as displacement in NMR state during evolution of the qubit-NMR coupled system. Clearly, the magnitude of displacement in NMR state depends on $\Omega_2$, further the displacement will be negative or positive depending on the qubit states $|\pm\rangle$. When $\Omega_2=0$, the hamiltonian changes to the optically driven qubit-NMR system \cite{auffeves}, where the effect of qubit-NMR coupling leads to second order modifications in qubit energy states as well as shift in frequency of NMR. Next we exploit this interaction for generating superposition of mesoscopic states of NMR.
\begin{figure}[h!]
\centering
\includegraphics[height=6 cm]{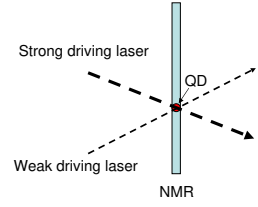}
\vspace{-0.1cm}
\caption{Schematic for a  nanomechanical resonator coupled with a two level quantum mechanical system (qubit). The qubit is driven by two external fields one field is strong and another is weak.}
\label{fig1}
\end{figure}

\section{Generating superposition of mesoscopic states of NMR}
\label{mesoscopic}
We consider initially qubit is in ground state $|g\rangle$ and NMR is in coherent state of phonons $|\alpha_0\rangle$, initial state can be written as
\begin{equation}
|\psi(0)\rangle=\frac{1}{\sqrt{2}}(|+\rangle-|-\rangle)|\alpha_0\rangle.
\end{equation}
In the presence of field $\Omega_1$, when $\Omega_2$ is switched on the phonon annihilation operator $a$ transforms to $A$, i. e. $a\rightarrow D(-i\Omega_2\lambda/\omega)aD(i\Omega_2\lambda/\omega)$ as a result the state of NMR changes as $|\alpha\rangle\rightarrow D(-i\Omega_2\lambda/\omega)|\alpha\rangle$, here $D$ is displacement operator. Similarly when $\Omega_2$ is switched off phonon annihilation operator transforms as $a\rightarrow D(i\Omega_2\lambda/\omega)aD(-i\Omega_2\lambda/\omega)$ and the state of NMR transforms as $|\alpha^{\prime}\rangle\rightarrow D(i\Omega_2\lambda/\omega)|\alpha^{\prime}\rangle$. As $\lambda$ has values $\pm1$ corresponding to qubit states $|\pm\rangle$, a small positive or negative displacement in NMR state is produced when $\Omega_2$ is switched on or off depending on the state of the qubit. In order to get constructive effect during one on-off cycle of $\Omega_2$, we work on a strategy that the time between switch on and off of field $\Omega_2$ should be such that the phonon field has changed its phase by $\pi$. Therefore we consider that $\Omega_1$ and $\Omega_2$ are switched on for half of the time period of NMR, i.e. for $t=0$ to $t=\pi/\omega$ and remain off for rest of the half. The state of the system at $t=\pi/\omega$ is given by
\begin{eqnarray}
\psi(\pi/\omega)=\frac{1}{\sqrt{2}}\left[e^{-i\phi}e^{-il_1\Re(\alpha_0)}|(\alpha_0-il_1)e^{-i(\omega-\Omega_1\eta^2)\frac{\pi}{\omega}}\rangle|+\rangle\right.\nonumber\\
\left.-e^{i\phi}e^{il_1\Re(\alpha_0)}|(\alpha_0+il_1)e^{-i(\omega+\Omega_1\eta^2)\frac{\pi}{\omega}}\rangle|-\rangle\right]\\
=\frac{1}{\sqrt{2}}\left[e^{-i\phi}e^{-il_1\Re(\alpha_0)}|(-\alpha_0+il_1)e^{il_2\pi}\rangle|+\rangle\right.\nonumber\\
\left.-e^{i\phi}e^{il_1\Re(\alpha_0)}|(-\alpha_0-il_1)e^{-il_2\pi}\rangle|-\rangle\right],
\end{eqnarray}
where $\Re$ represents the real part, $\phi=\Omega^{\prime}_1\pi/\omega$, $l_1=\Omega_2\eta/\omega$, and $l_2=\Omega_1\eta^2/\omega$. For time $t=\pi/\omega$ to $t=2\pi/\omega$, fields $\Omega_1$ and $\Omega_2$ are switched off, therefore the state of the system is given by
\begin{eqnarray}
\psi(2\pi/\omega)=\frac{1}{\sqrt{2}}\left[e^{-i\phi}e^{-il_1\Re[\alpha_0+(\alpha_0-il_1)e^{il_2\pi})]}|(\alpha_0-il_1)e^{il_2\pi}-il_1\rangle|+\rangle\right.\nonumber\\
\left.-e^{i\phi}e^{il_1\Re[\alpha_0+(\alpha_0+il_1)e^{-il_2\pi})]}|(\alpha_0+il_1)e^{-il_2\pi}+il_1\rangle|-\rangle\right].
\end{eqnarray}
If we measure the qubit in the ground state $|g\rangle$ or in the excited state $|e\rangle$, the state of the NMR is projected into a coherent superposition of mesoscopic state $|(\alpha_0-il_1)e^{il_2\pi}-il_1\rangle$ and $|(\alpha_0+il_1)e^{-il_2\pi}+il_1\rangle$. Clearly for small values of $l_1$ these states overlap considerably, therefore the effects of quantum interference become significant. If we assume that the qubit is measured in ground state $|g\rangle$ after switching off the fields, the projected state of NMR is given by
\begin{eqnarray}
\psi(2\pi/\omega)=\left[e^{-i\phi}e^{-il_1\Re[\alpha_0+(\alpha_0-il_1)e^{il_2\pi})]}|(\alpha_0-il_1)e^{il_2\pi}-il_1\rangle\right.\nonumber\\
\left.+e^{i\phi}e^{il_1\Re[\alpha_0+(\alpha_0+il_1)e^{-il_2\pi})]}|(\alpha_0+il_1)e^{-il_2\pi}+il_1\rangle\right].
\end{eqnarray}
 After sending one square pulse of each fields $\Omega_1$ and $\Omega_2$ simultaneously, which is on for half of time period of NMR and off for
the other half, the phonon field is displaced anticlockwise or clockwise along a circle in a random fashion. If we pass $N$ such pulse pairs and every time detect the qubit in
its ground state, the state of NMR will be equivalent to $n$ step random walk along a circle\cite{circle}. The state of NMR can be expressed as
\begin{eqnarray}
 |\psi_{ph}(n)\rangle=C\left[e^{-i\phi}\hat{O}(-l_1,-l_2)+e^{i\phi}\hat{O}(l_1,l_2)\right]^n|\alpha_0\rangle,
\end{eqnarray}
where $C$ is normalization constant and operator $\hat{O}(l_1,l_2)|\alpha\rangle=D(il_1)e^{-il_2\pi a^{\dagger}a}D(il_1)$.
We note that operators $\hat{O}(l_1,l_2)$ and $\hat{O}(-l_1,-l_2)$ commute each other $[\hat{O}(l_1,l_2),\hat{O}(-l_1,-l_2)]=0$ for real $l_1,~l_2$ and $\hat{O}(l_1,l_2)\hat{O}(-l_1,-l_2)|\alpha\rangle=|\alpha\rangle$. Therefore the state of NMR can be written as
\begin{eqnarray}
|\psi_{ph}(n)\rangle&=&
C\sum_{m=0}^{n} \left(
                    \begin{array}{c}
                      n \\
                      m \\
                    \end{array}
                  \right)\left[e^{-im\phi}\hat{O}^m\left(-l_1,-l_2\right)\times\right.~~~~~~~~\nonumber\\
&&\left.e^{i(n-m)\phi}
\hat{O}^{n-m}(l_1,l_2)\right]|\alpha_0\rangle,\nonumber\\
&=&C\sum_{m=0}^{n} \left(
                    \begin{array}{c}
                      n \\
                      m \\
                    \end{array}
                  \right)
e^{i(n-2m)\phi}\hat{O}^{n-2m}(l_1,l_2)|\alpha_0\rangle,\\
&=&C\sum_{m=0}^{N} \left(
                    \begin{array}{c}
                      n \\
                      m \\
                    \end{array}
                  \right)
e^{i(n-2m)\phi}e^{-i\theta_{n-2m}}|\alpha_{n-2m}\rangle, \label{final}
\end{eqnarray}
where we have recursive expressions $\theta_j=\theta_{j-1}+l_1\Re(\alpha_{j-1}+\alpha_j)$ and $\alpha_j=(\alpha_{j-1}+il_1)e^{-il_2\pi}+il_1$. Now expressing this result in coordinate
representation, we get expression for wavefunction for NMR $\psi_{ph}(n)=\langle x|\psi_{ph}(n)\rangle$
\begin{eqnarray}
\psi_{ph}(n)=C\sum_{m=0}^{n} \left(
                    \begin{array}{c}
                      n \\
                      m \\
                    \end{array}
                  \right)
e^{i(n-2m)\phi}e^{-i\theta_{n-2m}}\psi_{\alpha_{n-2m}}(x).
\label{nstate}
\end{eqnarray}
where $\psi_{\alpha_n}(x)=\pi^{-1/4}\exp[-(x-\sqrt{2}\Re(\alpha_n))^2/2+i\sqrt{2}\Im(\alpha_n)x-i\Re(\alpha_n)\Im(\alpha_n)]$; $\Im(\alpha_n)$ is imaginary part of $\alpha_n$.
\begin{figure}[h!]
\centering
\includegraphics[height=6 cm]{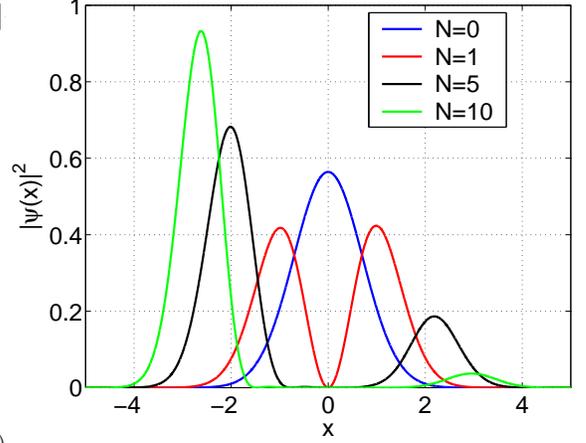}
\vspace{-0.1cm}
\caption{(Color online)Probability distribution $|\psi_{ph}(n)|^2$ for the amplitude of the nanomechaical resonator after passing different number of pulse pairs. The parameters are $l_1=0.1$, $l_2=0.01$, $\phi=9\pi/2$, and initial state as $|0\rangle$.}
\label{fig2}
\end{figure}
In Fig.\ref{fig2}, we have plotted the probability distribution of displacement of NMR using initial state $|\alpha_0\rangle=|0\rangle$ and typical values of $\phi$, $l_1$, $l_2$ for different values of $n$. The displacement of NMR  depends on $l_1$, $l_2$, $\phi$ and number of pulse pairs $n$. For $l_1=0.1$ and $l_2=0.01$, we calculate values of $\alpha_1=0.00314107591+0.199950656i$, $\alpha_5=0.0783720116+0.995810825i$, $\alpha_{10}=0.311558267+1.96710148i$with $\alpha_{-n}=\alpha^*_n$. The maximum expected value of displacement $x$ for a typical value of $n$ is given by $\langle x\rangle_n= \sqrt{2}\Re(\alpha_n)$. From Eq.(\ref{nstate}), it is clear that the final state of NMR is coherent superposition of $n+1$ coherent states, further there relative phases depends on $\phi$ and $\theta_s$. For small values of $l_1$ and $l_2$ these coherent states overlap considerably leading to dominating quantum interference effects. The unexpected displacement in the state of NMR is due to the constructive interference between the coherent states generated after passing $n$ pulse pairs. Here we must emphasize that the choice of $\phi$ is also critical for final displacement in NMR state. For $\phi$ equals to odd multiple of $\pi/2$, we get maximum displacement and for even multiple of $\pi/2$ the displacement is negligible. For $\phi$ equals to odd multiple of $\pi/4$ only one peak in probability distribution appears which has displacement in between maximum and minimum values. For $n=1$, we get two equal peaks symmetrically placed on positive and negatives sides around $x=0$. When value of $n$ increases, peak on negative side start dominating, in fact for $n\geq10$, the peak along positive side becomes negligible. The final state of NMR is equivalent to the quantum random walk defined by Aharnov et al.\cite{aharonov,pathak} along a circle.

Next, we discuss how one can generate a superposition state of two coherent states which are well separated in phase space. We consider that the strong driving field $\Omega_1$ is on for a few cycles of mechanical oscillations and the weak driving field is modified in a similar fashion as discussed above. We consider initial state of the system as $|g\rangle|\beta_0\rangle$ with $\beta_0=0$. Say, after $n$ cycles of mechanical oscillations, when $n$ pulses of $\Omega_2$ have been passed, $\Omega_1$ is switched off and we measure the state of qubit in ground state $|g\rangle$, the projected state of NMR is given by
\begin{eqnarray}
|\psi_{cat}(n)\rangle=K\left(e^{-i\phi^{\prime}}e^{i\theta^{\prime}_{-n}}|\beta_{-n}\rangle-e^{i\phi^{\prime}}e^{i\theta^{\prime}_{n}}|\beta_{n}\rangle\right),
\label{catstate}
\end{eqnarray}
where $K$ is normalization constant and $\phi^{\prime}=\Omega^{\prime}_12n\pi/\omega$, $\beta_j=(\beta_{j-1}+il_1)e^{-2il_2\pi}+il_1e^{-il_2\pi}$, $\theta_j=\theta_{j-1}+l_1\Re\left[\beta_{j-1}+(\beta_{j-1}+il_1)e^{-2il_2\pi}\right]$ with $\theta_0=0$. Clearly above method can be used to generate superposition of two mesoscopic states similar to Schrodinger cat states. In Fig.\ref{fig4} we have plotted the Wigner function for the state (\ref{catstate}).
\section{Decoherence}
\label{Sec:Decoherence}
\begin{figure}[h!]
\centering
\includegraphics[height=6 cm]{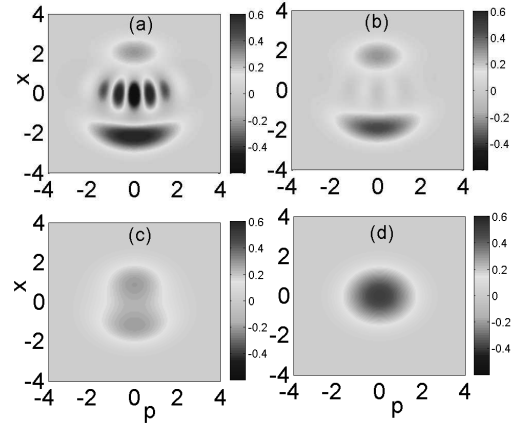}
\vspace{-0.1cm}
\caption{(Color online)Wigner function $W(x,p)$ of the generated superposition of mesoscopic states (\ref{nstate}) for $n=5$. In (a) $\xi=0$, in (b) $\xi=0.2$, in (c) $\xi=0.5$, and in (d) $\xi=1$, other parameters are same as in Fig.\ref{fig2}}
\label{fig3}
\end{figure}

In the Sec.\ref{mesoscopic} we have shown that the unexpected displacement in NMR state is produced due to quantum interference. Therefore, it is important to maintain coherence in the system during the generation of the superposition of mesoscopic states of NMR. For a nanomechanical resonator having high quality factor, we can neglect effects of decoherence in generated superposition state due to phonon mode damping\cite{decoherence}. In following we consider effects of decoherence in generated states due to spontaneous decay of qubit. During the generation of states (\ref{nstate}) and (\ref{catstate}), the qubit remains in dressed states $|+\rangle$ and $|-\rangle$. Using density matrix of qubit $\rho_q$ we evaluate the density matrix elements at time $t$, the diagonal elements remain constant $\langle +|\rho_q(t)|+\rangle=\langle -|\rho_q(t)|-\rangle=1/2$ and off diagonal elements decay as $\langle \pm|\rho_q(t)|\mp\rangle=\langle \pm|\rho_q(0)|\mp\rangle\exp(-3\Gamma t/4)$ \cite{scullybook}. We include effect of qubit decoherence in the generated state (\ref{nstate}) as follows. We calculate density matrix, $\rho_{ph}(n)$ for generated NMR state after $n$ pulse-pairs are passed using recursion relation
\begin{eqnarray}
\rho_{ph}(n)=C[\hat{O}(l_1,l_2)\rho_{ph}(n-1)\hat{O}(l_1,l_2)+\hat{O}(-l_1,-l_2)\nonumber\\
\rho_{ph}(n-1)\hat{O}(-l_1,-l_2)+e^{2i\phi}e^{-\xi}\hat{O}(l_1,l_2)\rho_{ph}(n-1)\nonumber\\
\hat{O}(-l_1,-l_2)
+e^{-2i\phi}e^{-\xi}\hat{O}(-l_1,-l_2)\rho_{ph}(n-1)\hat{O}(l_1,l_2)],
\label{deco}
\end{eqnarray}
with $\rho_{ph}(0)=|\alpha_0\rangle\langle\alpha_0|$; where $\xi=3\Gamma T/8$ and C is normalization constant. In Fig.\ref{fig3}, we plot Wigner function for state (\ref{deco}) for $n=5$ using different values of $\xi$. The Wigner function for the density matrix $\rho$ is defined as
\begin{equation}
W(x,p)=\frac{1}{\pi\hbar}\int e^{2ipy/\hbar}
\langle x-y|\rho|x+y\rangle dy.
\end{equation}
In Fig.3(a), for $\xi=0$, Wigner function shows two peaks at $x\approx\pm2$, one smaller peak at x=2 and one dominating peak at x=-2. The interference fringes are visible between these peaks which are direct signature of coherence. Wigner function acquires negative values in region between the peaks. The negative value of Wigner function clearly indicates nonclassical nature of the generated superposition state. Further squeezing in x quadrature\cite{squeezed} is also visible, which is also clear in Fig.\ref{fig2}. In Fig.\ref{fig3}(a) to (d), as the value of $\xi$ increases the interference diminishes progressively. In Fig.\ref{fig3}(d), for $\xi=1$, generated superposition state (\ref{nstate}) completely turn into a mixed state and the displacement becomes approximately zero.

\begin{figure}[h!]
\centering
\includegraphics[height=6 cm]{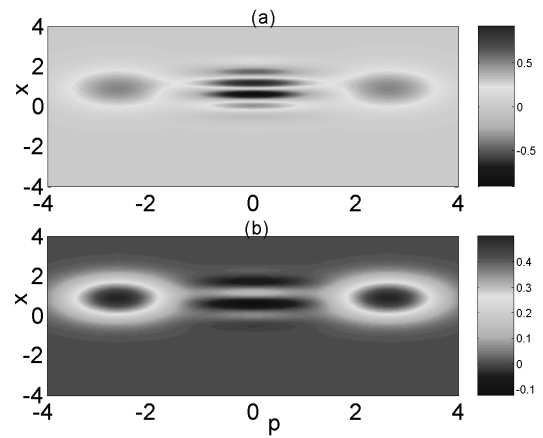}
\vspace{-0.1cm}
\caption{(Color online)Wigner function $W(x,p)$ of the generated superposition of two mesoscopic states (\ref{catstate}) for $n=10$. In (a) $\Gamma =0$, in (b) $3n\Gamma T/4=2$, other parameters are same as in Fig.\ref{fig2}}
\label{fig4}
\end{figure}

In Fig.\ref{fig4}(a), we plot Wigner function for state (\ref{catstate}) using $n=10$. In Fig.\ref{fig4}(b), Wigner function, for the same parameters used in Fig.\ref{fig4} (a) and for value of spontaneous decay rate such that $3n\Gamma T/4=2$. From Figs. \ref{fig3} \& \ref{fig4}, it is clear that the life time of qubit is a crucial factor for generating superposition of mesocopic states of NMR.

\section{Conclusions}
\label{Sec:Conclusions}
We have presented a scheme to generate superposition of mesoscpic states of a nanomechanical resonator by coupling with a two level quantum mechanical system. We have sown that as a result of quantum interference the displacement amplitude of the resonator could be exceptionally large we also find some squeezing in the position quadrature. We have considered decoherence effects due to spontaneous decay of two level quantum mechanical system. For $\xi=3\Gamma T/8=0.2$, we find small changes in Wigner function. This condition can be satisfied for resonator frequency $\omega\approx\Gamma$, thus a quantum dot coupled with a nanomechanical resonator of frequency $1GHz$ could be considered for possible experimental realization of our results. In NV center qubits, life time of qubit is very large and $\eta=g/\omega\sim10^{-3}$ can also be achieved. Therefore NV centers as a qubit using resonator of frequency $1MHz$ could be another system for potential realization.
\section{Acknowledgements}
This work was supported by DST SERB Fast track young scientist scheme SR/FTP/PS-122/2011.

\end{document}